\documentclass[%
 reprint,
 amsmath,amssymb,
 aps,dvipdfmx
]{revtex4-1}
\usepackage{graphicx}
\usepackage{dcolumn}
\usepackage{bm}
\usepackage{color}

\renewcommand{\a}{\alpha}
\renewcommand{\b}{\beta}

\newcommand{\bea}{\begin{eqnarray}}
\newcommand{\eea}{\end{eqnarray}}
\newcommand{\f}[2]{\frac{#1}{#2}}
\newcommand{\eq}{&=&}
\newcommand{\nn}{\nonumber \\ }
\newcommand{\ve}{\varepsilon}
\renewcommand{\d}{\delta}
\newcommand{\area}{\int_{-\infty}^\infty }

\newcommand{\p}{\partial}
\newcommand{\pp}[2]{\f{\p #1}{\p #2}}

\newcommand{\sref}[1]{Eq. (\ref{#1})}

\newcommand{\citeauthorname}[2]{{#1} {#2}}
\newcommand{\citebook}[4]{{#1} {\it #2} ({#3}, {#4}).}
\newcommand{\citepaper}[4]{{#1} {#3} ({#4}).}
\begin{document}

\preprint{APS/123-QED}

\title{Maximizing and Minimizing Investment Concentration \\ with Constraints of Budget and Investment Risk}

\author{Takashi Shinzato}
\email{takashi.shinzato@r.hit-u.ac.jp}
 \affiliation{
Mori Arinori Center for Higher Education and Global Mobility,
Hitotsubashi University, 
Tokyo, 1868601, Japan.}

\date{\today}

\begin{abstract}
In this paper,  
as a first step in examining the properties of 
a feasible portfolio subset that is characterized by 
budget and risk constraints, we assess 
the maximum and minimum of the investment concentration using replica analysis.
To do this, we apply an analytical approach of statistical mechanics. 
We note that the
optimization problem considered in this paper is 
the dual problem of the portfolio optimization problem discussed in 
the literature, and we verify that these optimal solutions are also 
 dual. We also present numerical experiments, in which we use the method of
steepest descent that is based on Lagrange's method of undetermined multipliers, 
and we compare the numerical results to those obtained by replica analysis in order to assess the effectiveness of our proposed approach.
\begin{description}
\item[PACS number(s)]
{89.65.Gh}, {89.90.+n}, {02.50.-r}
\end{description}
\end{abstract}
\maketitle

\section{Introduction}
The portfolio optimization problem is 
one of the most important research topics in the area of 
mathematical finance, and 
it is well known that
the investment risk can be reduced by diversifying 
assets in accordance with the 
knowledge obtained from the optimal solutions to this problem \cite{Bodie,Luenberger}.
The pioneering research on this topic was 
reported by Markowitz in 1952 \cite{Markowitz1952,Markowitz1959}, and it is still an active area of research \cite{Konno,Rockafellar}.
Several recent studies have considered 
investment models that use the
analytical approaches developed in 
cross-disciplinary fields, such as 
replica analysis, belief propagation methods, and using the distribution of the
eigenvalues of random matrices \cite{Ciliberti1,Ciliberti2,Kondor,Pafka,Shinzato-SA2015,Shinzato-BP2015,VH,Shinzato-qw-fixed2016}.
For instance, 
Ciliberti et al. \cite{Ciliberti1,Ciliberti2} 
used replica analysis in the limit of absolute zero temperature to examine the minimal investment risk per asset when using
the absolute deviation model or the expected shortfall model.
Kondor et al. \cite{Kondor} quantified the sensitivity to noise 
for several risk functions, including the in-sample risk, the out-sample risk, and the predicted risk. Moreover, 
Pafka et al. \cite{Pafka} investigated the relationship between the number of investment periods and the value of assets, as well as various 
investment risks such as the predicted risk and the practical risk. Shinzato \cite{Shinzato-SA2015} used replica analysis to show 
that for the mean-variance model, 
the minimal investment risk and its concentration 
are 
self-averaging. Furthermore,
Shinzato et al. \cite{Shinzato-BP2015} developed an 
algorithm to solve for 
the optimal portfolio when using the mean-variance model and the absolute deviation model
and a belief propagation method, and they
proved the Konno-Yamazaki conjecture for a quenched disordered system. 
Varga-Haszonits et al. \cite{VH} used replica analysis to investigate
the minimal investment risk and the efficient frontier for the mean-variance model under 
budget and return constraints. In addition, 
Shinzato \cite{Shinzato-qw-fixed2016} used replica analysis to investigate 
the minimal investment risk for the mean-variance model with budget and investment concentration constraints.

Of the studies discussed above, {
the minimal investment risk for a mean-variance model with a number of constraints 
was analyzed only in Ref. \cite{Shinzato-qw-fixed2016} as
a natural extension of the
mean-variance model with a budget constraint 
considered in Ref. \cite{Shinzato-SA2015};}
it turns out that the dual problem 
is implied in these portfolio optimization problems. 
In order to better
understand these optimization problems, 
we use the dual structure to analyze them. However, 
in the various investigations of this problem that have used analytical approaches that were developed in 
cross-disciplinary fields (including replica analysis and an approach based on using the distribution of the
eigenvalues of random matrices), 
there are few studies 
that analyze the potential of an investment system that 
proactively employs a dual structure and the dual problem.
As a first step in 
discussing the mathematical framework of a dual structure, 
our aim in this paper is to solve the dual problem of the 
portfolio optimization problem 
\cite{Shinzato-qw-fixed2016}
 and to clarify the dual structure of these optimization problems.

This paper is organized as follows: 
in section II, 
we state the dual problem of the portfolio optimization problem with 
budget and investment concentration constraints, as discussed in 
Ref. \cite{Shinzato-qw-fixed2016}.
In section \ref{sec3}, 
we use replica analysis to investigate this dual problem. 
In section \ref{sec4}, we compare the results of the replica analysis to those estimated by numerical experiments and 
evaluate the effectiveness of our proposed method. In section 5, we present our conclusions and discuss areas of future work.

\section{Model Setting}
As in Refs. \cite{Shinzato-SA2015,Shinzato-BP2015,Shinzato-qw-fixed2016}, we consider a stable investment market in which there is no regulation of 
short selling and in which there are $N$ assets.
A portfolio of asset $i(=1,\cdots,N)$ is notated as $w_i$, and
a portfolio of $N$ assets is notated as 
$\vec{w}=(w_1,w_2,\cdots,w_N)^{\rm T}\in{\bf R}^N$. We will use the 
notation ${\rm T}$ to mean the transpose of a vector or a matrix.
For simplicity, we assume that
short selling is not regulated, and 
we note that $w_i$ is nonnegative.
We assume $p$ scenarios, and
the return rate of asset $i$ in scenario $\mu(=1,\cdots,p)$ is 
$\bar{x}_{i\mu}$, 
where the return rates are independently distributed 
with a mean $E_X[\bar{x}_{i\mu}]$ and unit
variance. We will consider the feasible portfolio subset $W(\kappa)$, 
which is subject to the following constraints on the budget and risk constraint:
\bea
N\eq\sum_{i=1}^Nw_i,\label{eq1}\\
N\kappa\ve \eq\f{1}{2}
\sum_{\mu=1}^p
\left(\f{1}{\sqrt{N}}
\sum_{i=1}^Nw_i\left(
\bar{x}_{i\mu}-E_X[\bar{x}_{i\mu}]
\right)
\right)^2,
\label{eq2}
\eea
where \sref{eq1} is the budget constraint used in Refs. {\cite{Shinzato-SA2015,Shinzato-qw-fixed2016}},
\sref{eq2} is a risk constraint, and 
$\ve$ is the minimal investment risk $\ve=\f{\a-1}{2}$. 
Note that \sref{eq2} 
implies that 
the investment risk for $N$ assets is $\kappa(\ge1)$ times the minimal 
investment risk $N\ve$. 
We will call $\kappa$ the risk coefficient, and 
the scenario ratio is defined as $\a=p/N$. In addition, the
modified return rate $x_{i\mu}$ is defined as
$x_{i\mu}=\bar{x}_{i\mu}-E_X[\bar{x}_{i\mu}]$, and so the feasible 
portfolio subset $W(\kappa)\subseteq{\bf R}^N$ can be rewritten as follows:
\bea
\label{eq3}
W(\kappa)\eq
\left\{
\vec{w}\in{\bf R}^N\left|N=\vec{w}^{\rm 
T}\vec{e},N\kappa\ve=\f{1}{2}\sum_{\mu=1}^p\left(\f{\vec{w}^{\rm T}\vec{x}_\mu}{\sqrt{N}}\right)^2
\right.
\right\},\nn
\eea
where the unit vector $\vec{e}=(1,\cdots,1)^{\rm T}\in{\bf R}^N$, and 
the modified return rate vector is 
$\vec{x}_\mu=(x_{1\mu},x_{2\mu},\cdots,x_{N\mu})^{\rm T}\in{\bf R}^N$. That is, 
the
Wishart matrix $XX^{\rm T}\in{\bf R}^{N\times N}$
defined by the
modified return rate matrix $X=\left\{\f{x_{i\mu}}{\sqrt{N}}\right\}\in{\bf 
R}^{N\times p}$ is the metric of the Mahalanobis distance (or, more accurately, half the squared Mahalanobis distance), $\f{1}{2}\vec{w}^{\rm T}XX^{\rm T}\vec{w}$,
which is constant. We need to 
examine the 
portfolios included in the feasible subset $W(\kappa)$ 
in order to investigate the properties of the investment market. We will use 
the following statistic, which has been used previously in the literature (e.g., \cite{Shinzato-SA2015}): 
\bea
q_w\eq\f{1}{N}
\sum_{i=1}^Nw_i^2.\label{eq4}
\eea 
For instance, when $\kappa=1$,  
the optimal solution is unique; 
when $\kappa>1$, the feasible subset 
$W(\kappa)$ is not empty, 
and if we can determine the range investment concentrations, then 
we can determine the number of portfolios in that subset.

Finally, we note that a previous study \cite{Shinzato-SA2015} examined
the optimal solution 
that minimizes the investment risk in \sref{eq2}
under the budget constraint in \sref{eq1}, and it also analyzed
the minimal investment risk.
A different study \cite{Shinzato-qw-fixed2016} examined 
the optimal solution that minimizes the investment risk in 
\sref{eq2} 
under the budget constraint in \sref{eq1}
and the investment concentration constraint in \sref{eq4}, and again, it analyzed 
the minimal investment risk.
We note that this study \cite{Shinzato-qw-fixed2016}, which discusses the 
portfolio optimization problem with two constraints, is a natural extension of the previous study \cite{Shinzato-SA2015}, which considered only a single constraint. 
In this paper, we 
interchange 
the investment concentration constraint 
and the object function (the investment risk) to consider the dual of the problem considered in Ref. \cite{Shinzato-qw-fixed2016}.

\section{Replica analysis\label{sec3}}
In this section, we use replica analysis \cite{Mezard,Nishimori} to investigate the 
optimization problem discussed above.
The Hamiltonian in this investment system is
\bea
{\cal H}(\vec{w})\eq\f{1}{2}\sum_{i=1}^Nw_i^2.
\label{eq5}
\eea
Following the approach of statistical mechanics, 
the partition function $Z(\kappa,X)$ of 
the inverse temperature $\b$ is 
\bea
Z(\kappa,X)\eq
\area d\vec{w}P(\vec{w}|\kappa,X)e^{\b{\cal H}(\vec{w})},\label{eq5}\\
P(\vec{w}|\kappa,X)\eq\d\left(
\sum_{i=1}^Nw_i-N
\right)\nn
&&
\d\left(
N\kappa\ve-
\f{1}{2}
\sum_{\mu=1}^p
\left(\f{\vec{w}^{\rm T}\vec{x}_\mu}{\sqrt{N}}
\right)^2
\right),
\eea
where $X=\left\{\f{x_{i\mu}}{\sqrt{N}}\right\}\in{\bf 
R}^{N\times p}$ is the return rate {matrix.}
From this, the maximum and minimum of the investment concentration, 
$q_{w,\max}$ and $q_{w,\min}$, respectively, can be derived using the following formula: 
\bea
q_{w,\max}\eq
\mathop{\max}_{\vec{w}\in W(\kappa)}
\left\{
\f{1}{N}
\sum_{i=1}^Nw_i^2
\right\}\nn
\eq\lim_{\b\to\infty}\f{2}{N}
\pp{}{\b}\log Z(\kappa,X),\label{eq9}\\
q_{w,\min}\eq
\mathop{\min}_{\vec{w}\in W(\kappa)}
\left\{
\f{1}{N}
\sum_{i=1}^Nw_i^2
\right\}\nn
\eq\lim_{\b\to-\infty}\f{2}{N}
\pp{}{\b}\log Z(\kappa,X).\label{eq10}
\eea
In order to assess the bounds of the investment concentration, we use 
the unified viewpoint approach of statistical mechanics, 
although we do not use the Boltzmann factor, which is widely used 
in the literature of statistical mechanics. 
Since this representation maintains the mathematical structure of this model, 
we can analyze both bounds within large limits of the inverse temperature $\b$. In addition, 
in order to examine the typical behavior of this investment system, 
we need to evaluate the typical maximum and minimum investment concentrations.
That is, 
we must 
rigorously average the right-hand side in 
\sref{eq9} and 
\sref{eq10} over the return rate of assets.

In a way similar to that used in previous studies 
\cite{Shinzato-SA2015,Shinzato-qw-fixed2016},
we used replica analysis and the ansatz of the replica symmetry solution (see 
Appendix \ref{sec-appA} for details), as follows:

\bea
\label{eq11}
\phi\eq\lim_{N\to\infty}\f{1}{N}E_X[\log Z(\kappa,X)]\nn
\eq\mathop{\rm Extr}_{k,\theta,\chi_w,q_w,\tilde{\chi}_w,\tilde{q}_w}\left\{
-k+\kappa\theta\ve+\f{1}{2}(\chi_w+q_w)(\tilde{\chi}_w-\tilde{q}_w)\right.\nn
&&+\f{q_w\tilde{q}_w}{2}
+\f{\b}{2}(\chi_w+q_w)+\f{k^2}{2\tilde{\chi}_w}-\f{1}{2}\log\tilde{\chi}_w
+\f{\tilde{q}_w}{2\tilde{\chi}_w}\nn
&&\left.-\f{\a}{2}\log(1+\theta\chi_w)-\f{\a\theta q_w}{2(1+\theta\chi_w)}\right\},
\eea
where $\mathop{\rm Extr}_mf(m)$ 
is the extremum of function $f(m)$ with respect to $m$ , and the 
replica symmetry solution is evaluated at 
$a,b=1,2,\cdots,n$, as follows:
\bea
\label{eq11-1}
q_{wab}\eq
\left\{
\begin{array}{ll}
\chi_w+q_w&a=b\\
q_w&a\ne b
\end{array}
\right.,\\
\tilde{q}_{wab}\eq
\left\{
\begin{array}{ll}
\tilde{\chi}_w-\tilde{q}_w&a=b\\
-\tilde{q}_w&a\ne b
\end{array}
\right.,\\
k_a\eq k,\\
\theta_a\eq\theta,
\label{eq14-1}
\eea
where $k$ is the auxiliary variable with respect to 
\sref{eq1}, and $\theta$ is
 the auxiliary variable with respect to 
\sref{eq2}. 
From this, the extremum conditions in \sref{eq11} are derived as follows:
\bea
k\eq\tilde{\chi}_w,\\
\chi_w\eq\f{1}{\tilde{\chi}_w},\\
q_w\eq1+\f{\tilde{q}_w}{\tilde{\chi}_w^2},\\
\tilde{\chi}_w+\b\eq\f{\a\theta}{1+\theta\chi_w},\\
\tilde{q}_w\eq\f{\a\theta^2q_w}{(1+\theta\chi_w)^2},\\
\f{\kappa(\a-1)}{2}\eq\f{\a\chi_w}{2(1+\theta\chi_w)}+
\f{\a q_w}{2(1+\theta\chi_w)^2}.
\eea
In order to obtain the maximum and minimum, we need to take the 
limit as $|\b|\to\infty$; 
we use the results presented in Refs.
\cite{Shinzato-SA2015,Shinzato-qw-fixed2016}. Then, we assume 
$\theta\chi_w\sim O(1)$ and 
$\f{\b}{\theta}\sim O(1)$, and so we obtain
\bea
\theta\chi_w\eq\f{1\pm\sqrt{\a-\f{\a}{\kappa}}}{\a-1},\label{eq17}\\
\f{\b}{\theta}\eq\pm\f{(\a-1)^2\sqrt{\a-\f{\a}{\kappa}}}{2\a-\f{\a}{\kappa}\pm(\a+1)\sqrt{\a-\f{\a}{\kappa}}}.\label{eq18}
\eea
From these, we then obtain
\bea
\chi_w\eq\pm\f{(\a-1)\sqrt{\a-\f{\a}{\kappa}}}{\b\left(\a\pm\sqrt{\a-\f{\a}{\kappa}}\right)},\\
q_w\eq\f{\left(\sqrt{\a\kappa}\pm\sqrt{\kappa-1}\right)^2}{\a-1},\label{eq19}
\eea
where, from the seventh term in \sref{eq11}, we have
$-\f{1}{2}\log\tilde{\chi}_w=\f{1}{2}\log\chi_w$, 
since $\chi_w>0$. 
Note that if $\b>0$, the $\chi_w$, and $q_w$ are both positive, and if 
$\b<0$, they are both negative.
Moreover, from Eqs. (\ref{eq9}), (\ref{eq10}), and 
(\ref{eq11}), we obtain
\bea
\lim_{N\to\infty}2\pp{}{\b}
\left\{\f{1}{N}E_X[\log Z(\kappa,X)]\right\}
\eq
2\pp{\phi}{\b}\nn
\eq\chi_w+q_w,
\eea
is obtained. 
Since $\chi_w$ is close to $0$, then when $|\b|\to\infty$, we obtain
\bea
q_{w,\max}\eq\f{\left(\sqrt{\a\kappa}+\sqrt{\kappa-1}\right)^2}{\a-1},\label{eq21}\\
q_{w,\min}\eq\f{\left(\sqrt{\a\kappa}-\sqrt{\kappa-1}\right)^2}{\a-1}.\label{eq22}
\eea

Four points should be noted here. First,  
both bounds of the investment concentration are 
consistent when $\kappa=1$, and so 
$q_{w,\max}=q_{w,\min}=\f{\a}{\a-1}$. Second, 
the maximum investment concentration $q_{w,\max}$ has no upper bound, while 
the minimum investment concentration $q_{w,\min}$ has a lower bound at $\kappa=\f{\a}{\a-1}$,
and so $q_{w,\min}=1$. Third, 
the optimization problem discussed in the literature is the dual problem of
the one considered in the present work.
When $\tau=q_{w,\max}$, 
 $\kappa=\f{(\a+1)\tau-1-2\sqrt{\a\tau(\tau-1)}}{\a-1}$, and so 
the investment risk per asset 
$\ve'=\kappa\ve$ is calculated as follows:
\bea
\ve'\eq\f{\a\tau+\tau-1-2\sqrt{\a\tau(\tau-1)}}{2}.
\eea
We note that this coincides with the minimal investment risk per asset obtained in our 
previous {studies \cite{Shinzato-qw-fixed2016,Tada}}. That is, 
the portfolio in $W(\kappa)$ that maximizes the 
investment concentration corresponds to the portfolio in \bea
R(\tau)\eq
\left\{\vec{w}\in{\bf R}^N
\left|\vec{w}^{\rm T}\vec{e}=N,
{\vec{w}^{\rm T}\vec{w}}=N\tau
\right.
\right\}
\label{eq31}
\eea
that minimizes the investment risk. If 
$\tau=q_{w,\min}$ and 
$\kappa=\f{(\a+1)\tau-1+2\sqrt{\a\tau(\tau-1)}}{\a-1}$, then
the investment risk per asset 
$\ve''=\kappa\ve$ is 
\bea
\ve''\eq\f{\a\tau+\tau-1+2\sqrt{\a\tau(\tau-1)}}{2};
\eea
this 
corresponds to the maximal investment risk per asset found in Refs.
\cite{Shinzato-qw-fixed2016,Tada}; that is, 
the portfolio in $W(\kappa)$ that minimizes 
the investment concentration corresponds to 
the portfolio in $R(\tau)$ in \sref{eq31} that maximizes 
the investment risk.

The fourth point considers the annealed disordered system for this investing strategy 
(for a detailed explanation of annealed and
quenched disordered systems, see \cite{Shinzato-SA2015}).
From our previous studies {\cite{Shinzato-SA2015,Shinzato-BP2015}}, 
the minimal expected investment risk per asset of an annealed disordered 
system is $\ve^{\rm OR}=\f{\a}{2}$, and so the risk constraint in \sref{eq2} is 
replaced by
\bea
N\kappa\ve^{\rm OR}\eq\f{1}{2}\sum_{\mu=1}^p
E_X\left[
\left(\f{\vec{w}^{\rm T}\vec{x}_\mu}{\sqrt{N}}\right)^2
\right]\nn
\eq
\f{\a}{2}\sum_{i=1}^Nw_{i}^2,
\eea
where $E_X[x_{i\mu}x_{j\mu}]=\d_{ij}$.
From this, the feasible portfolio subset of the annealed disordered system 
is calculated as follows:
\bea
W^{\rm OR}(\kappa)\eq
\left\{
\vec{w}\in{\bf R}^N\left|N=\vec{w}^{\rm 
T}\vec{e},\f{N\kappa\a}{2}=\f{\a}{2}\sum_{i=1}^Nw_{i}^2
\right.
\right\}.\nn
\eea
Thus, the maximum and minimum of the investment concentration 
$q_{w}^{\rm OR}$ are the same:
\bea
q_{w}^{\rm OR}\eq\kappa.
\eea
The feasible portfolio subset 
$W(\kappa)$ in 
\sref{eq3} is determined by the portfolio $\vec{w}$ for which half of the squared 
Mahalanobis distance is consistent; note that the metric of the Mahalanobis distance 
is defined by the Wishart matrix $XX^{\rm T}$, which is derived from 
the return rate matrix $X=\left\{\f{x_{i\mu}}{\sqrt{N}}\right\}\in{\bf R}^{N\times p}$.
However, in general, since this feasible portfolio subset 
$W(\kappa)$ is not isotropic, 
the portfolio closest to the origin (which minimizes the 
investment concentration) and 
the one farthest from the origin (which maximizes the 
investment concentration) are uniquely determined. 
However, since the feasible portfolio subset of the annealed disordered system $W^{\rm 
OR}(\kappa)$ is isotropic, 
this implies that the maximum and minimum investment concertation are the same.

\section{Numerical Experiments\label{sec4}}

In order to evaluate the effectiveness of our proposed approach, 
we numerically assess the maximum and minimum investment concentration, 
$q_{w,\max}$ and $q_{w,\min}$, respectively, and compare
 the results with those obtained by replica analysis. We replace the feasible portfolio subset 
 $W(\kappa)$ in \sref{eq3} with 
constraint conditions using 
Lagrange's method of undetermined multipliers, 
 and the object function of Lagrange's method $L(\vec{w},k,\theta)$ is 
defined as follows:
\bea
L(\vec{w},k,\theta)\eq\f{1}{2}\vec{w}^{\rm T}\vec{w}+k(N-\vec{e}^{\rm T}\vec{w})
+\theta
\left(\f{1}{2}\vec{w}^{\rm T}J\vec{w}-N\kappa\ve\right),
\nn
\eea
where $k,\theta$ are the auxiliary variables, and the $i,j$th component 
 of the Wishart matrix $J(=XX^{\rm 
T})=\left\{J_{ij}\right\}\in{\bf R}^{N\times N}$ is
\bea
J_{ij}\eq\f{1}{N}\sum_{\mu=1}^px_{i\mu}x_{j\mu}.
\eea
It is necessary to evaluate  
 the optimal solution of the object function of Lagrange's method, $L(\vec{w},k,\theta)$,
in order to determine the maximum and minimum of investment concentration. We used the following method of steepest descent: 
\bea
\vec{w}^{s+1}\eq\vec{w}^s-\eta_w\pp{L(\vec{w},k,\theta)}{\vec{w}},\\
k^{s+1}\eq k^s+\eta_k\pp{L(\vec{w},k,\theta)}{k},\\
\theta^{s+1}\eq \theta^s+\eta_\theta\pp{L(\vec{w},k,\theta)}{\theta},
\eea
where, at step $s$, the portfolio is $\vec{w}^s=(w_{1}^s,w_{2}^s,\cdots,w_{N}^s)^{\rm 
T}\in{\bf R}^N$ and the auxiliary variables are $k^s,\theta^s\in{\bf R}$; also,
$\vec{w}^0=\vec{e}$ and $k^0=\theta^0=1$.
When $\eta_k,\eta_\theta,\eta_w>0$, we can determine the minimum, 
and when $\eta_k,\eta_\theta,\eta_w<0$,
we can determine the maximum.
The stopping condition is that  
$\Delta=\sum_{i=1}^N|w_{i}^s-w_{i}^{s+1}|+|k^s-k^{s+1}|+|\theta^s-\theta^{s+1}|$
is less than $\d$.

From this, we obtain the
$M$ return rate matrices, $X^1,\cdots,X^M$, where the 
$m$th return rate matrix is $X^m=\left\{\f{x^m_{i\mu}}{\sqrt{N}}\right\}\in{\bf 
R}^{N\times p}$, 
with respect to the risk coefficient 
$\kappa$, using $q_{w,\max}(\kappa,X^m)$ and $q_{w,\min}(\kappa,X^m)$, as
estimated using the algorithm given above. These 
are calculated as follows:
\bea
\label{eq39}q_{w,\max}(\kappa)\eq\f{1}{M}\sum_{m=1}^Mq_{w,\max}(\kappa,X^m),\\
\label{eq40}q_{w,\min}(\kappa)\eq\f{1}{M}\sum_{m=1}^Mq_{w,\min}(\kappa,X^m),
\eea
where the return rate of asset $i$, $x_{i\mu}^m$, is 
independently and identically distributed with zero
mean and unit variance.

\begin{figure}[tbh]
\begin{center}
\includegraphics[width=0.9\hsize,angle=0]{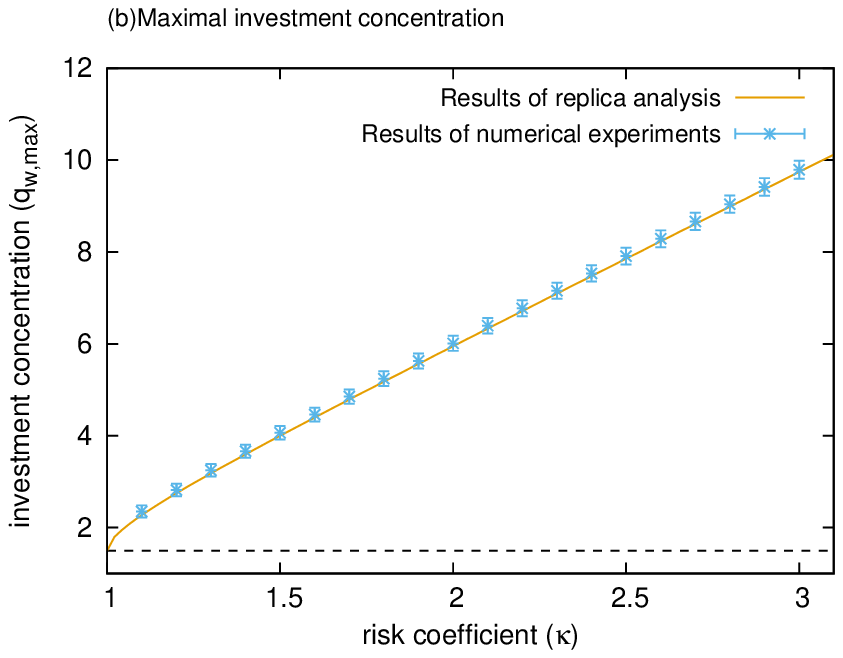}
\caption{
\label{Fig1}
Comparison of the maximal investment concentration obtained by the replica 
 analysis to that obtained in the numerical experiments; $\a=p/N=3$.
The horizontal axis shows the risk coefficient $\kappa$, and the vertical axis shows 
the minimal investment concentration $q_{w,\min}$.
The solid line (orange) shows the results of the replica analysis, the asterisks with 
 error bars (blue) show the results of the numerical simulation, and 
the dashed line shows the investment concentration at $\kappa=1$, that 
 is, $\f{\a}{\a-1}$.
}
\includegraphics[width=0.9\hsize,angle=0]{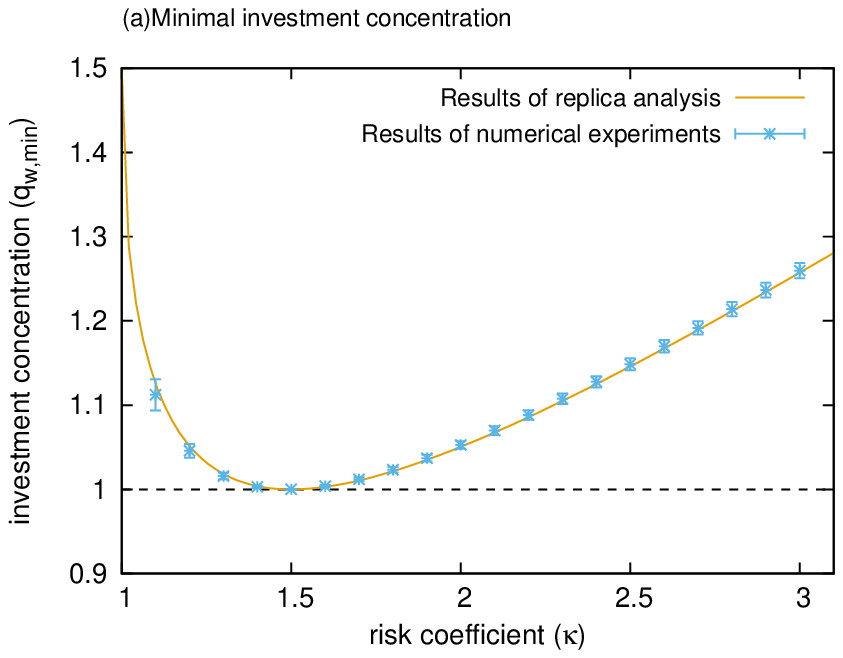}
\caption{
\label{Fig2}
Comparison of the minimal investment concentration obtained by the replica 
 analysis to that obtained in the numerical experiments; $\a=p/N=3$.
The horizontal axis shows the risk coefficient $\kappa$, and the vertical axis shows 
the maximal investment concentration $q_{w,\max}$.
The solid line (orange) shows the results of the replica analysis, the asterisks with 
 error bars (blue) show the results of the numerical simulation, and the 
 dashed line shows the minimal investment concentration, that 
 is, unity.
}
\end{center}
\end{figure}

We performed numerical experiments with the following settings:
$N=1000, p=3000$, $\a=p/N=3$, and $M=10$.
When seeking the minimum, we used $\d=10^{-5}$,
{$\eta_k=10^{-1},\eta_\theta=10^{-5}$, and $\eta_w=10^{-1}$,} and when 
seeking the maximum, we used {$\eta_k=-10^{-1},\eta_\theta=-10^{-5}$, and $\eta_w=-10^{-1}$.} The results of the replica analysis and numerical 
experiments are shown in Figs. \ref{Fig1} and \ref{Fig2}.
The horizontal axis shows the investment concentration $q_w$, and 
the vertical axis shows the risk coefficient $\kappa$.
Solid lines are the results of the replica analysis 
(Eqs. (\ref{eq21}) and (\ref{eq22})) and 
the asterisks with error bars 
are the results of the numerical simulation 
(Eqs. (\ref{eq39}) and (\ref{eq40})).
The figures show that 
the results of the replica analysis are consistent with those of the
numerical simulation, 
and so 
we can use replica analysis to accurately analyze the
portfolio optimization problem.

\section{Conclusion and Future work}

In the present study, 
we used replica analysis, which was developed for cross-disciplinary research,
to analyze 
the duality problem of the portfolio optimization problem with several 
constraint conditions, which has been considered 
in our previous studies {\cite{Shinzato-SA2015,VH,Shinzato-qw-fixed2016}}. 
We determined a feasible portfolio that maximizes the investment concentration subject to budget and risk constraints, and 
one that minimizes the investment concentration.
We applied a canonical ensemble analysis to a large, complicated system 
with respect to this optimization problem with several restrictions.
From a unified viewpoint, 
we were able to derive the maximum and minimum investment concentrations 
from the subset of feasible portfolios.
The portfolio optimization problem considered in this paper is the dual 
of the optimization problem discussed in our previous {study \cite{Shinzato-qw-fixed2016}}, 
and we verified that the optimal solutions possess the duality structure. 
In the numerical experiments, 
we used the method of steepest descent that is based on Lagrange's method of undetermined multipliers, and
we compared the numerical and theoretical results to verify our proposed approach.

In this and our previous {studies \cite{Shinzato-SA2015,Shinzato-qw-fixed2016,Tada}}, 
we analyzed a portfolio optimization problem subject to several constraints. In the future, we intend to further 
examine the
complicated relationship between this and the dual problem in more general 
situations.
In addition, we intend to examine the effects of regulating short selling.

\section*{Acknowledgements}
The author appreciates the fruitful comments of K. Kobayashi and D. Tada. 
This work was supported in part 
by Grant-in-Aid No. 15K20999; the President Project for Young Scientists at Akita Prefectural
University; research project No. 50 of the National Institute of Informatics, Japan; research project No. 5 of
the Japan Institute of Life Insurance; research project
of the Institute of Economic Research Foundation at Kyoto University; research project No. 1414 of the Zengin
Foundation for Studies in Economics and Finance; research project No. 2068 of the Institute of Statistical
Mathematics; research project No. 2 of the Kampo
Foundation; and research project of the Mitsubishi UFJ Trust Scholarship Foundation.

\appendix
\section{Replica analysis\label{sec-appA}}
In this appendix, we explain {the replica analysis} used in the present paper.
As in Ref. \cite{Shinzato-SA2015},
$E_X[Z^n(\kappa,X)]$, $n\in{\bf Z}$, is defined as follows:
\bea
&&E_X[Z^n(\kappa,X)]\nn
\eq
\f{1}{(2\pi)^{\f{Nn}{2}+pn}}
\area \prod_{a=1}^nd\vec{w}_ad\vec{u}_ad\vec{v}_a
E_X\left[
\exp\left(\f{\b}{2}\sum_{i=1}^N\sum_{a=1}^nw_{ia}^2\right.
\right.\nn
&&+\sum_{a=1}^nk_a\left(\sum_{i=1}^Nw_{ia}-N\right)
+\sum_{a=1}^n\theta_a\left(N\kappa\ve-\f{1}{2}\sum_{\mu=1}^pv_{\mu a}^2\right)\nn
&&
\left.\left.
+i\sum_{\mu=1}^p\sum_{a=1}^nu_{\mu a}
\left(v_{\mu a}-\f{1}{\sqrt{N}}
\sum_{i=1}^Nx_{i\mu}w_{ia}
\right)
\right)\right].
\eea
In the thermodynamic limit of the number of assets $N$, we obtain
\bea
&&\lim_{N\to\infty}\f{1}{N}\log E_X[Z^n(\kappa,X)]\nn
\eq\f{\b}{2}{\rm Tr}Q_w-\vec{k}^{\rm 
T}\vec{e}+\kappa\ve\vec{\theta}^{\rm T}\vec{e}+\f{1}{2}{\rm 
Tr}Q_w\tilde{Q}_w+\f{1}{2}\vec{k}^{\rm T}\tilde{Q}_w^{-1}\vec{k}\nn
&&-\f{1}{2}\log\det|\tilde{Q}_w|-\f{\a}{2}\log\det\left|
\begin{array}{rr}
Q_w&-iI\\
-iI&\Theta
\end{array}
\right|,
\label{eq-a2}
\eea
where we have the order parameters 
\bea
q_{wab}\eq\f{1}{N}\sum_{i=1}^Nw_{ia}w_{ib},
\eea
and the conjugate parameters $\tilde{q}_{wab}$. Here,
$k_a$ is the auxiliary variable with respect to \sref{eq1}, and 
$\theta_a$ is the auxiliary variable with respect to
\sref{eq2}. In addition, 
in \sref{eq-a2}, 
$Q_w=\left\{q_{wab}\right\}\in{\bf R}^{n\times n}$, 
$\tilde{Q}_w=\left\{\tilde{q}_{wab}\right\}\in{\bf R}^{n\times n}$, 
$\vec{k}=(k_1,\cdots,k_n)^{\rm T}\in{\bf R}^n$, 
$\vec{\theta}=(\theta_1,\cdots,\theta_n)^{\rm T}\in{\bf R}^n$, 
$\vec{e}=(1,\cdots,1)^{\rm T}\in{\bf R}^n$, and 
$\Theta={\rm diag}\left\{\theta_1,\theta_2,\cdots,\theta_n\right\}\in{\bf R}^{n\times n}$. If we substitute the replica symmetry solutions from Eqs. (11) to (14) into \sref{eq-a2},
we obtain
\sref{eq11}.

\end{document}